\documentclass[titlepage]{amsart}
\usepackage{graphicx}
\theoremstyle{definition}

\theoremstyle{remark}

\newcommand{\norm}[1]{\left\Vert#1\right\Vert}
\newcommand{\abs}[1]{\left\vert#1\right\vert}

\newcommand{\bA}{\mathbf{A}}

\newcommand{\bC}{\mathbf{C}}

\newcommand{\bE}{\mathbf{E}}
\newcommand{\bU}{\mathbf{U}}
\newcommand{\tbR}{\tilde{\mathbf{R}}}
\newcommand{\tbH}{\tilde{\mathbf{H}}}

\newcommand{\tq}{\tilde{q}}
\newcommand{\tQ}{\tilde{Q}}
\newcommand{\tPhi}{\tilde{\Phi}}
\newcommand{\tphi}{\tilde{\phi}}
\newcommand{\tx}{\tilde{x}}

\newcommand{\tzeta}{\tilde{\zeta}}
\newcommand{\el}{\ell}

\newcommand{\cderii}[2]{\ensuremath{\frac{{d^2} #1}{{d} {#2}^2}}}
\newcommand{\rr}{\mathbf{r}}
\newcommand{\vv}{\mathbf{v}}

\newcommand{\bSig}{{\boldsymbol{\Sigma}}}

\newcommand{\Erf}{\mathrm{Erf}}

\newcommand{\uu}{\underline{u}}

\begin{document}

\title[]{Iterative Determination of Distributions by  the Monte Carlo Method in Problems with an External Source }%
\author{Mih\'aly Makai}%
\email{makai@reak.bme.hu,makai.mihaly@energia.mta.hu}%
\author{Zolt\'an Szatm\'ary}%
\email{szatmary@reak.bme.hu}%
\address{BME NTI, H-1111 Budapest, Muegyetem rkp. 3-9,}%

\date{\today}%

\begin{abstract}
In the Monte Carlo (MC) method statistical noise is usually present. Statistical noise may become dominant  in the calculation of a distribution, usually by iteration, but is less Important  in calculating  integrals. The subject of the present work is the role of statistical noise in  iterations involving stochastic simulation (MC method). Convergence is checked by  comparing two consecutive solutions in the iteration. The statistical noise may randomize or pervert the convergence.  We study the probability of the convergence, and the correct estimation of the variance in a simplified model problem. We study the statistical properties of the solution to a deterministic problem with a stochastic source obtained from a stochastic calculation. There are iteration strategies resulting in non-convergence, or randomly stopped iteration.
\end{abstract}
\maketitle
\vspace{1cm}
\verb"Accepted for publication in Nuclear Science and Engineering, May 10, 2013"
\verb"Fax: 36-1-4631954"
\verb"e-mail: makai@reak.bme.hu, makai.mihaly@energia.mta.hu"
\verb"szatmary@reak.bme.hu"
\verb"Address: BME NTI, H-1111 Budapest, Muegyetem rkp. 3-9"
\pagebreak
\section{Introduction}
\label{sec:Intro}
%
%

Neutron diffusion must be one of the first applications of the Monte Carlo (MC) method\cite{Ulam}. Since then, several monographes{\footnote{The references have been subjectively selected.}} appeared discussing various aspects of the MC method~\cite{Gelbard}, \cite{Lux} have appeared.

Today the MC method is the most popular numerical modeling method. Most of the applications  aim at determining a discretized distribution function: spatial power distribution, burnup distribution, temperature distribution.

The recent literature does not pay due attention to the statistical aspects of the MC method being applied to determine distributions. In some cases\cite{Gudow} the parallelism with deterministic methods is overemphasized. Statistical noise is an inherent property of the MC method. Compare the convergence check in a deterministic and a stochastic iteration. Assume our end is to determine the neutron distribution at 1000 points. In a deterministic method, it suffices to compare the relative differences in two consecutive iteration steps to check the convergence. When the absolute difference is below a given limit, the iteration has converged. On the other side, the distribution determined by the MC method is a random function described by its mean $\eta_i, i=1,\dots,N_c$ and standard deviation $\sigma_i$ at each point. The standard deviation is usually approximated by $\sigma=1/\sqrt{N_h}$ where $N_h$ is the number of histories simulated to obtain the estimate of $\eta_i$. The error limit $\varepsilon=0.01$ assures only low accuracy, although it requires $N_h=10^4$ histories. If $\sigma_i=0.01$ for all $i$, the probability of a $3\sigma$ fluctuation is $0.0027$, the number of average outliers is $27$, their position will be random and the procedure may never converge even if the convergence criterion is as large as $\varepsilon=0.02$.

The statistical noise plays an important role in any stochastic iteration when we intend to determine a vector or a distribution function. The reason is that the convergence criterion should be applied to each component of the vector or distribution.

Our goal is to study the statistics of distributions determined by the MC method in an iteration. To this end we study two simplified model problems. In either case we seek the discretized solution, that we call flux, of a linear problem with a random source. The statistics of the source are known: the source is normally distributed at every point, the expectation values and the variances are given. In problem I. the random source is a given random distribution and we seek the expectation value and variance of the discretized flux. The solution is determined by an iteration. In problem II. the random source is recalculated in each iteration step, assuming that expectation values and variances do not change during the iteration. We will show that in general the variance increases with the number $N_c$ of points. We show that in the finite difference (FD) approximation the MC result is unbiased, and in problem II. the variance is considerably larger than the theoretical value, in some cases the variance does not diminish with the progress of the iteration.

Our notation is as follows. Random variables are labeled by tilde: $Q_i$ is deterministic,  but $\tQ_i$ is a stochastic variable. Vectors are underlined: $\underline{Q}=(Q_1,Q_2,\dots,Q_{n})$, and
$\underline{\tQ}=(\tQ_1,\tQ_2,\dots,\tQ_{n})$. The expectation value of $\tQ_i$ is written as $E(\tQ_i)$, the variance is $D^2(\tQ_i)$. As to statistics, we follow the notation in Ref. \cite{Papoulis}.

We summarize the basic ideas applied throughout the present work as follows\cite{LA}:
\begin{enumerate}
  \item We consider a source problem and the source is considered random but given. The mean value and variance of the source are assumed to be given{\footnote{Actually, the source may have been determined by another MC code.}}.
  \item The problem under consideration is deterministic in the following sense: the cross sections and geometry are given.
  \item We seek a discretized solution in $N_c>>1$ phase space cells.
  \item We solve the problem by computer simulation: generate random numbers, from the random numbers we build a history and, from $N_h>>1$ histories, determine the mean value and the variance for each phase space cell.
  \item The obtained solution gives a characterization of the neutron flux. We assume the determined flux values to be identically distributed, the probability density function of the flux being the normal distribution, variances and expectation values may vary at different cells.
  \item We assume $N_h$ to be large enough for the neutron flux to follow the normal distribution.
\end{enumerate}

The above described calculation is one of the large number of typical applications of the MC  method in reactor physics. The suggested simplifications serve the feasibility of the analysis.

Another important aspect of the approximate solutions is the discretization. Human beings and  computers are capable of carrying out only finitely many computations. Consequently, we are able to work only with finite models. A finite model is described as follows.  Let the possible coordinates and velocities take at most $N_r$ and $N_v$ values, respectively.  If solutions are determined in $N_t$ time intervals, a solution is represented by $N_r N_v N_t$ numbers. We consider two solutions identical if they take the same values on every interval. The discretized solution is a vector
\begin{equation}\label{Fund-4}
    \phi(n_r,n_v,n_t),\;\; 1\le n_r\le N_r;\;1\le n_v\le N_v;\; 1\le n_t \le N_t.
\end{equation}
 Let the tolerance limit of the approximate solution be $\varepsilon$. If there are  iteration steps $\el$ and $\el+1$ such that
\begin{equation}\label{Fund-5}
    |\phi_{\el+1}(n_r,n_v,n_t)- \phi_\el(n_r,n_v,n_t)|\le \varepsilon,
\end{equation}
 or
 \begin{equation}\label{Fund-5a}
   \frac{ |\phi_{\el+1}(n_r,n_v,n_t)- \phi_\el(n_r,n_v,n_t)|}{\phi_\el(n_r,n_v,n_t)}\le \varepsilon
\end{equation}
 holds for every $n_r,n_v,n_t$, then the iteration is called converged.

In Annex A, we describe the root finding method~\cite{Kim} shortly, and determine the mean and variance of the solution. In Annex B, we present a detailed analysis: the one-dimensional finite difference method is simple enough to carry out the mentioned calculations. We show that the application of the MC method requires special attention because some iteration methods give false results.


 It is known that the solution to the neutron transport equation for a given source term is unique\cite{Csase Zw}, and, in principle, can be determined by deterministic methods with any desired finite precision. The nature of the solution provided by the Monte Carlo method is completely different. First of all, there is no unique Monte Carlo solution, even if we run a given code on a given computer. In a given calculation, the Monte Carlo solution is based on a finite number of numerically simulated experiments and such a computation yields a random function $\tilde{\Phi}_{MC}(\rr,\vv,t)$ in the form of a stochastic process. If the Monte Carlo method is unbiased the Central Limit Theorem (CLT) claims this stochastic process to converge to the true solution in the stochastic sense.

The present day practice of Monte Carlo (MC) calculations includes several applications where MC is utilized in iterations.

  To avoid mathematical complications, we consider a simplified model problem in which the source $\underline{\tQ}=(\tQ_1,\dots,\tQ_{N_c})$ is random and given. We solve the equation by the Monte Carlo method using stochastic root finding. As to the source, its elements are assumed to be statistically independent and identically distributed. The expectation value and variance of each element is $Q_i$ and $\sigma^2$, respectively. The distribution function of $\tQ_i$ is $N(Q_i,\sigma^2)$. We wish to solve equation \eqref{RF-1} for $\underline{\tPhi}$ using stochastic root finding described in Appendix A.

\section{Applying stochastic root finding}
First we separate the source $\tQ$ and the dependent variable that we call flux $\tPhi$ into a deterministic and a random part:
\begin{equation}\label{Asrf-1}
     \tQ_i = Q_i + \tq_i,\;\;i=1,\dots,N_c;
\end{equation}
from the flux we separate the expectation value $\Phi_i=E(\tPhi_i)$:
\begin{equation}\label{Asrf-2}
   \tPhi_i = \Phi_i + \tphi_i \;\;i=1,\dots,N_c.
\end{equation}
We have to solve the following equation:
\begin{equation}\label{Asrf-3}
    \underline{\tPhi}= \bA \underline{\tPhi} + \underline{\tQ}.
\end{equation}
Here $\bA$ is a deterministic matrix. Taking the expectation value of Eq. \eqref{Asrf-3}, we immediately see that
\begin{equation}\label{Asrf-4}
    \underline{\Phi}= \bA \underline{\Phi} + \underline{Q}.
\end{equation}
Subtract Eq. \eqref{Asrf-4} from Eq. \eqref{Asrf-3} to obtain
\begin{equation}\label{Asrf-5}
    \underline{\tphi}= \bA \underline{\tphi} + \underline{\tq}.
\end{equation}
The next step in stochastic root finding is to set up an iteration to determine $\underline{\tphi}$. This is trivial in our case: the iteration goes as
\begin{equation}\label{Asrf-6}
    \underline{\tphi}_{\el+1} = \bA \underline{\tphi}_\el + \underline{\tq},\el=0,1,2,\dots
\end{equation}
with $\underline{\tphi}_0=\underline{\tq}$. Or, using recursion \eqref{Asrf-6} repeatedly,
\begin{equation}\label{Asrf-7}
    \underline{\tphi}_{\el+1} = \bA^\el \underline{\tphi}_0 + \sum_{\el'=0}^{\el-1} \bA^{\el'} \underline{\tq},
\end{equation}
and when $\norm{\bA}<1$ this converges to
\begin{equation}\label{Asrf-8}
    \underline{\tphi}= (\bE - \bA)^{-1} \underline{\tq}.
\end{equation}
The next step is to investigate the statistics.

As to the statistics of the converged error vector, its expectation value is zero, and its covariance matrix is
\begin{equation}\label{Asrf-9}
    E\{\underline{\tphi} \underline{\tphi}^T\}= (\bE - \bA)^{-1} E\{ \underline{\tq} \underline{\tq}^T\} (\bE - \bA)^{-1},
\end{equation}
in which superscript $T$ refers to the transposed vector. Assume that $\bA$ is an invertible matrix, then its eigenvectors $\uu_k$ are determined from the following equation:
\begin{equation}\label{Asrf-10}
    \bA \uu_k= \lambda_k \uu_k, k=1,2,\dots,N_c,
\end{equation}
and the matrix
\[
\bU=(\uu_1,\uu_2,\dots,\uu_{Nc})
\]
formed from the column vectors $\uu_k, k=1,2,\dots,N_c$ is invertible. When matrix $\bA$ is symmetric we have $\bA^T=\bA^{-1}$, where superscript $T$ stands for transposed matrix.
Since the components of vector $\underline{\tq}$ are statistically independent and their variances are all equal to $\sigma^2$, the covariance matrix is
\begin{equation}\label{Con-10}
     E\{\underline{\tphi} \underline{\tphi}^T\}= \sigma^2 (\bE - \bA)^{-2}= \sigma^2 \bU \mathbf{L} \bU^T,
\end{equation}
 and matrix $\mathbf{L}$ is diagonal with entries
\begin{equation}\label{Con-10a}
    L_{kk} = \frac{1}{(1-\lambda_k)^2}.
\end{equation}

We conclude that the iteration converges to $(\bE - \bA)^{-1}E\{\underline{\tq}\} $. The variance of the converged solution depends on the eigenvalues of the matrix of the iteration. If the convergence is slow then there must be an eigenvalue close to one, say $\lambda=1-\delta$, $\delta<<1$. Then, the variance of the solution is proportional to $\sigma^2/\delta^2$. The variance monotonically increases with the number of points $N_c$ because
\[
(\bU \mathbf{L} \bU^T)_i = \sum_{j=1}^{N_c} \frac{u_{ij}^2}{(1-\lambda_j)^2}>0\;\;\text{ for all }i.
\]

Since $E\{\tq_i\}=0$ for all $i$, the probability of convergence is one. The variance of the solutions is larger than the variance of the source and when $N_c\to \infty$ then
\begin{equation}\label{Con-11}
     E\{\underline{\tphi} \underline{\tphi}^T\}
\end{equation}
also tends to infinity. The practical consequence is that when solving a large, linear problem with a given random source, the variances of the local fluxes grow with the number of cells. Note that the solution method \eqref{Asrf-6} is deterministic: the iteration involves only the deterministic matrix $\bA$. Thus it corresponds to the problem $g(\Phi)=0$ in Appendix A.

We conclude the present Section by adding two remarks:
\begin{itemize}
  \item The iteration \eqref{Asrf-7} applies the deterministic matrix $\bA$ to the source and to the vector $\underline{\tPhi}_0$.
  \item In an iteration, the source term may be recalculated with various frequencies. This iteration, when the random source is given, is called iteration of type I. Later we investigate another iteration strategy, when the source is recalculated in each iteration step from an already converged Monte Carlo method. That iteration strategy is called iteration of type II.
\end{itemize}

\section{Convergence and statistics of stochastic vectors}
\label{sec:Conv}
The solution $\underline{\tPhi}$ of equation \eqref{Asrf-3} is a random vector of $N_c$ elements. The iteration \eqref{Asrf-6} is converged ~\cite{Gudow},\cite{LA} when
\begin{equation}\label{Cssv-1}
    \abs{\frac{\tPhi_{i,\el} - \tPhi_{i,\el+1}}{\tPhi_{i,\el}}}<\varepsilon
\end{equation}
for every $1\le i \le N_c$. Since the convergence criterion is applied to stochastic quantities, the criterion can be met only by a given probability.

 Below we determine the probability of the convergence of a stochastic iteration. The key features of the stochastic iteration are:
\begin{enumerate}
  \item The average number of points $N_p$ in a random walk called history.
  \item The number of histories $N_h$.
  \item The number $N_c$ of mesh cells in the phase space.
  \item The probability distribution of a given tally in a given cell.
  \item The probability that the convergence criteria are met.
\end{enumerate}
$N_p$ is usually chosen by practical considerations. When $N_p$ is small, the variance $\sigma_0$ within a history may be large. To monitor the suitability of $N_p$, the mean score, the relative error or other indicators can be used. For example the MCNP program is well equipped with tests helping the user assess the appropriateness of the chosen parameters.

Stochastic modeling is based on the CLT: the number of stochastic events should be large enough to achieve a reasonably small variance. Note, that the standard deviation is the product of two components: $\sigma_0$ and $1/\sqrt{N_h}$. A small variance does not exclude arbitrarily large local statistical fluctuations.

The number of cells $N_c$ depends on the physics of the problem under consideration, in general $N_c=N_v N_r N_t$ where $N_v$-the number of velocity cells, $N_r$-the number of spatial cells, $N_t$-the number of time steps. In practical calculations $N_c$ is between $10^3$ and $10^5$.

When $N_h>50$, the central limit theorem is applicable and the probability distribution  of a tally is practically the normal distribution. It is not sure if the probability distributions at different cells, i.e. $\tPhi_{i_1}$ and $\tPhi_{i_2}$ are statistically independent. The basic problem is that in the iteration to solve Eq. \eqref{Asrf-3} fluxes of various iteration steps mix. The correlation can be estimated from the tallies.

The probability of convergence depends on two conditions. The first is that the actual value of any given tally in two consecutive iteration steps should not differ by more than the given convergence limit $\varepsilon$. In a stochastic method that criterion is met with a given probability:
\begin{equation}\label{On-1}
   \left(\int_{-\varepsilon/2}^{+\varepsilon/2} N[\eta,\sigma(\xi)]d\xi\right)^2.
\end{equation}
The second condition requires  the estimated solution to be closer to the expectation value than $\varepsilon$ in each phase cell. As we have seen, the iteration is unbiased because the expectation value of the error is zero. The MC solution is the mean value estimated from $N_h$ histories:
\begin{equation}\label{Cstv-1}
    \tPhi_i^{MC}= \frac{\sum_{k=1}^{N_h} \tPhi_{i}^{(k)}}{N_h}
\end{equation}
where $\tPhi_{i}^{(k)}$ stands for the solution in the $k$-th history and phase cell $i$. Under the above fixed conditions, the mean value $\tPhi_i^{MC}$ is normally distributed with  variance $\sigma_0/\sqrt{N_h}$, where $\sigma_0$ is the variance within a given history.
Finally, we have to estimate the probability of the local random events
\begin{equation}\label{Cstv-2}
    \mathfrak{C}_i=\Phi_i-\varepsilon/2 \le\tPhi_i\le \Phi_i+ \varepsilon/2\;\;1\le i\le N_c ,
\end{equation}
and the global event $\mathfrak{C}_g$ when every $\mathfrak{C}_i$ is true, which is the following direct product (logical "and")
\begin{equation}\label{Cstv-3}
    \mathfrak{C}_g= \mathfrak{C}_1 \mathfrak{C}_2\dots \mathfrak{C}_{N_c}.
\end{equation}
Under the stipulated assumptions the probability of the event $\mathfrak{C}_i$ is independent of $i$:
\begin{equation}\label{Cstv-4}
    q=P(\mathfrak{C}_i)= \Erf\left(\frac{\varepsilon}{\sigma \sqrt{2}}\right),
\end{equation}
and because the probabilities are independent of index $i$, we obtain
\begin{equation}\label{Cstv-5}
    P_{conv}= P(\mathfrak{C}_g)=q^{N_c}.
\end{equation}
On the other hand, according to Eq. \eqref{Con-10} the variance of $\tPhi_i,\;i=1,\dots,N_c$ is  the variance $\sigma^2$ of the source term multiplied by an expression monotonically increasing with $N_c$.

The MC solutions are often used as  reference solution of a test problem where the solution must be determined at a large number $N_c>>1$ of phase-space points, and the convergence limit $\varepsilon$ is small. A phase space point is said to be an outlier if its value is beyond the expectation value plus or minus $\varepsilon$.

$N_h$ and $N_p$ are assumed to have been chosen so that local convergence be almost sure.  An outlier point destructs the convergence. Eq. \eqref{Cstv-5} gives the probability of convergence . In a statistical iteration, that probability is never one but may be close to one. Convergence of the iteration depends on two parameters: $q$ and $N_c$;  $q$ depending on $\varepsilon$ and $N_h$, see equation \eqref{On-1}.

 When $1-q$ is small, we may assume that $1-q$ is  inversely proportional to the number of phase-space points, and $N_c>>1$:
\begin{equation}\label{Asbe-2}
    1-q= \frac{B}{N_c}.
\end{equation}
The positive $B$ parameter depends on $\varepsilon$ and may be determined as $B=N_c (1-q)$. Now we write Eq. \eqref{Cstv-5} as
\begin{equation}\label{Asbe-3}
    P_{conv}= q^{N_c}= (1- (1-q))^{N_c}= \left(1-\frac{B}{N_c}\right)^{N_c}.
\end{equation}
Assume that the problem under consideration is large, thus $N_c\to \infty$; and we seek the exact solution $q\to 1$. Using the limit
\begin{equation}\label{Asbe-4}
    \lim_{n\to\infty} \left(1-\frac{a}{n}\right)^n= e^{-a},
\end{equation}
for any $a$, we find that
\begin{equation}\label{Asbe-5}
   \lim_{N_c\to\infty} P_{conv}=  e^{-B}.
\end{equation}
When $B=0$, the probability of convergence is 1, i.e,   the convergence is almost sure, but with increasing $B$ the convergence becomes less probable.

When $N_c=1000, q=1-0.001$ and $B=N_c(1-q)=1000\cdot 0.001=1$; then $P_{conv}=1/e=0.368$.
Once we have fixed the acceptable probability of convergence, the $N_h, N_p, \sigma_0$ values set a limit on the accuracy $\varepsilon$ of the calculation.

$B$ can be estimated by a numerical procedure, $q$ is estimated from Eq. \eqref{Cstv-4}; and using Eq. \eqref{On-1}, we find
\begin{equation}\label{Asbe-6}
    q =  \Erf\left( \frac{\sqrt{N_h}\varepsilon}{\sqrt{2}\sigma_0} \right).
\end{equation}
After fixing the desired convergence probability $P_{conv}$, we determine $q$ from Eq. \eqref{Cstv-4}. We determine the needed number of histories $N_h$ from Eq. \eqref{Asbe-6} when $\varepsilon$ is given . An example: when $N_c=10000, P_{conv}=0.9, \sigma_0=1$ and $\varepsilon=10^{-5}$, we need $N_h=7.73\;10^{11}$ histories.

\section{Practice of Monte Carlo calculations}

\begin{figure}[p]
  \includegraphics[width=7cm]{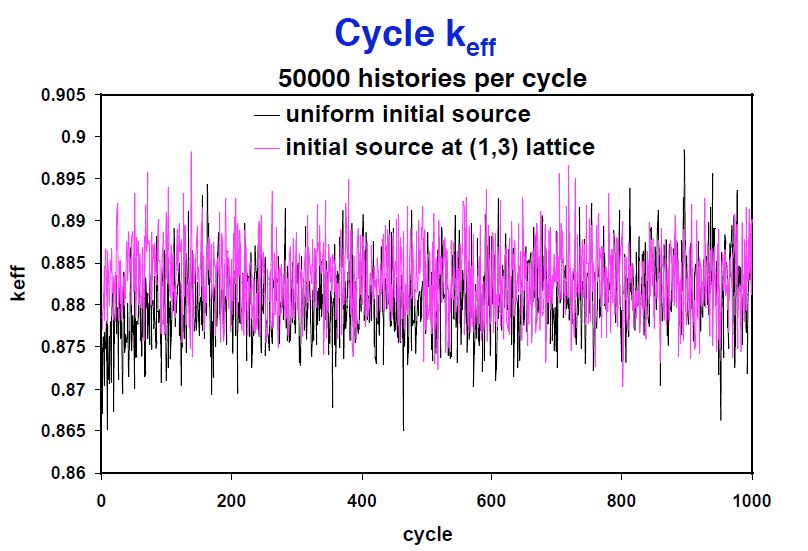}\\
  \caption{Progress of $k_{eff}$ iteration in MCNP}\label{F-MCNP1}
\end{figure}

In practice, Monte Carlo calculations use no convergence criteria but apparently are complacent about the applicability of the CLT. The first example is a figure from a MCNP report~\cite{LA}. In Figure \ref{F-MCNP1}, we see the progress of the $k_{eff}$ eigenvalue in a MCNP run, see Ref. \cite{LA}[p. 7-26]. The error of the $k_{eff}$ diminishes slowly and statistical fluctuations dominate the figure that clearly shows  the lack of convergence caused by the statistical fluctuations.

 For new fuel types, the desire to reduce margins has initiated  more and more sophisticated calculation methods. In \cite{MCNPCOBRA}, an excellent MC code (MCNP) and an also excellent fuel assembly code (COBRA-TF) have been combined. The authors mention that they use local convergence criteria for the fuel temperature, their criteria are $0.6\;\%$ in a benchmark calculation. The authors complain about the 180 hours computation time although they show in their Figure 6 only nine iteration steps, probably the maximal iteration step has been limited to 10, as can be deduced from their Figure  10, where the relative error of power is about 5 \%.

In the MC method discussed by us, the flux in a given cell is a random variable, and in iteration steps $\el$ and $\el+1$ their values would converge if they have tended to the same, definite random value. Note that this condition corresponds to the convergence criterium applied in deterministic iterations. Such a test is used in Ref. \cite{LA}[Lecture 8: Eigenvalue Calculation, II.]

In practical applications only the mean value estimate is checked  instead of the above mentioned test, see \cite{MCNPCOBRA},\cite{MCNPTRIPOLI}. This is a far looser "convergence" condition than any of the above mentioned definitions.

In Ref. \cite{MCNPTRIPOLI}, the iteration has not been mentioned, although in their Figure 6, nine iteration steps have been shown. On page 20 of their communication they make the following remark: \emph{"The convergence test is another issue. Currently the fuel temperature values in a pin from two successive iterations are tested whether they are within a certain range of deviation. It needs further study whether the coolant temperature and coolant density should also be included included in a convergence test." } It may be a bad piece of news that in an iteration the convergence test should be applied to all quantities modified in the iteration.

The MC method is used also in GEN IV. projects, in analysis~\cite{Reiss} of nonlinear problems and as reference \cite{Mar} as well. The mentioned applications have been selected subjectively.

\section{Concluding remarks}
We investigated the stochastic determination of a vector in an iteration. To this end we studied a simple model problem: a discretized linear equation with given random source, see Eq. \eqref{Asrf-4}. We assumed the source to be normally distributed, i.e. the probability density is $N(1,\sigma)$.

First we studied the exact solution, which gives a stochastic vector that we determined by an iteration. The solution is unbiased as the expectation value of the error vector is zero. The variance of the solution vector, see Eq. \eqref{Con-10}, is proportional to the variance of the source term, which is multiplied by a term monotonically increasing with the number of phase space points $N_c$. This means, the larger is $N_c$, the larger is the variance, and it becomes more probable that the statistical noise prohibits the point-wise convergence.

When matrix $\bA$ is deterministic, formulas have been derived to estimate the statistical noise. To cope with the random nature of the source, the calculation is organized in the same manner as in the MC programs: first a random walk is created with $N_p$ points and the random walks are repeated in $N_h$ histories. The relationships among $N_p, N_h, \varepsilon$, and $P_{conv}$ are given in Section \ref{sec:Conv}.

The stochastic root finding~\cite{Kim} offers a unified treatment of the deterministic and MC solutions of a given problem. The random walk is regarded as a stochastic approximation to the deterministic matrix $\bA$, as described in Appendix A. As a result, a new source term appears in the iteration, see Eq. \eqref{AppB-R-4}. That term can be regarded as another source term, this time generated by the stochastic method. When $N_h\to\infty$ that term vanishes.

It is a natural desire to analyze the statistics of the iteration, and the solution more closely. To this end Appendix  B discusses equation \eqref{Asrf-4} for a homogeneous material, diffusion theory, and in the finite difference approximation. We derived the statistics of the solution including the correlation matrix. The simple formalism has allowed not only for obtaining closed expressions for the covariance matrix of the theoretical solution, see Eq. \eqref{a.10-a}, but also studying two iteration strategies. In the first one the random source is fixed, in the second one the source is recalculated in each iteration step. In the latter case correlation is observed between distant points, see Fig. \ref{F-SzXX}, and the empirical variance differs from the theoretical, Fig. \ref{F-SzAA} and Fig. \ref{F-SzBB}. There are situations where the variance randomly fluctuates around a constant value, independently of the progress of the iteration.

In most practical reactor physics applications the problem is nonlinear because of feed-back effects (e.g. temperature, burnup). In accident analysis even the input may vary depending on the calculated dependent variables (flux or temperature). In those cases it should be demonstrated that the Monte Carlo results are adequate and should be explained why.

Our conclusion is that the very nature of the stochastic method does not set any limit to the problems solvable by it. On a 64 bit processor, the number of statistically independent random numbers is estimated as $10^{14}$ a number presumably sufficiently large for practical problems{\footnote{In MCNP5 the cycle length of the random number generator is close to $10^{19}$.}}.

There are, however, a few practical issues. It is a question if the required number of arithmetic operations can be finished in a reasonable time. Having $N_c$ phase space points, the number of tallies grows linearly with $N_c$, the number of calls of the random number generator is also linear in $N_c$. The variance of the solution monotonically increases with $N_c$, see Eqs. \eqref{Con-10} and \eqref{Con-10a}. It is true, however, that the stochastic iteration converges only with a given probability, and when that probability is too low the principally correct iteration may never end. This problem is more pronounced when $N_c$ is large enough for the outliers to slow down or prevent the convergence.

We propose to attach the following parameters to a solution obtained by an iteration in which Monte Carlo method is involved:
\begin{itemize}
  \item the target accuracy of the iteration $\varepsilon$;
  \item the probability of the convergence $P_{conv}$.
\end{itemize}
Those parameters are useful in judging the merit of the applied method. There are several works analyzing the Monte Carlo technics. The authors have hardly seen any report or paper mentioning that the iteration is a stochastic process and the convergence is attained only with a given probability. The probability of convergence of a given algorithm is not determined and disclosed. When $N_c$ tends to infinity, the probability of convergence tends to zero, even in the simple case that we have investigated. Correlations may worsen the situation. We did not discuss the probability of convergence in an eigenvalue problem but because of the imbedded iterations the problem is likely more severe than in the case discussed by us.

In reactor related applications, the statistical aspects of the problem are often disregarded and replaced by a general reference to the CLT. This is rather comfortable because more  attention may be given to the numerical or algorithmic side of the problem~\cite{Gudow}.

What is the situation in a deterministic iteration? Can we solve there arbitrary large problems? The answer is definitely no. The reason is twofold. The toughest obstacle is the finite accuracy of the computer arithmetic. Detailed investigations have shown that the computer addition is not an associative operation and the result depends on the sequence of the additions~\cite{Numer}. In meteorology, the number of unknowns is in the order of $10^6$. So it is possible to overcome the first difficulty. The second problem is the time of the calculation. In most numerical methods (finite element, finite difference, nodal method) the system matrix is sparse thus effective iterative solution methods have been elaborated. If we assume that in one iteration the number of arithmetic operations is approximately $N_c^2$, we conclude that the present day computer capacity allows for solving problems with $N_c\sim 10^6$. It is a question, however, what will be the probability of convergence.


\section{Appendix A: Stochastic root finding}
\label{sec:Srf}
Iteration is used in numerical methods, for example to find a root of a nonlinear equation. When at least one stochastic variable is involved, the iteration is called stochastic root finding\cite{Kim}. In this formalism, the problem we wish to solve  is formulated as an implicite equation{\footnote{Variable $\underline{g}$ has nothing to do with variable $g$ in Eqs. \eqref{Cstv-3} or \eqref{Cstv-5}.}}:
\[ \underline{g}(\underline{x})=0, \]
where the unknowns have been collected in vector $\underline{x}$ and we have as many equations as unknowns, therefore $\underline{g}$ is a vector-vector function. Note that $\underline{g}$ is deterministic. Notwithstanding, when there is a random parameter in $\underline{g}$ the solution becomes random. This is the case when the source is random:
\begin{equation}\label{RF-1}
    \underline{g}(\underline{\tPhi})\equiv (\bE-\bA)\underline{\tPhi} -\underline{\tQ}=0,
\end{equation}
where the source term $\underline{\tQ}$ is a random vector, $\bE$ is the identity matrix of order $N_c$, $\bA$ is a given deterministic matrix of order $N_c$, and we wish to determine the random vector $\underline{\tPhi}$. $\underline{g}=(g_1,\dots,g_{N_c})$ maps a vector of component $N_c$ into another vector of component $N_c$.

We have to set forth a list of key notations. Consider the problem $\underline{g}(\underline{x})=0$, where $\underline{g}$ and $\underline{x}$ are vectors ($N_c$ tuples). The norm of $x$ is given by $\norm{\underline{x}}=\sqrt{x_1^2 + x_2^2+\dots+x_{N_c}^2}$.

If $\underline{x}, \underline{y}\in \mathbb{R}^{N_c}$, by $\underline{x}\ge \underline{y}$ we mean that each element of $\underline{x}$ is not less than the corresponding element of $\underline{y}$. The sequence of random vectors $\{\underline{\tx}_n\}$ is said to \emph{almost surely converge} to a random vector $\{\underline{\tx}\}$, written as $\{\underline{\tx}_n\to\underline{\tx}\}$ a.s., if probability of $\lim_{n\to\infty}\{\underline{\tx}_n =\underline{\tx}\}$ equals to one. The sequence of random vectors $\{\underline{\tx}_n\}$ is said to  \emph{converge} to a random vector $\{\underline{\tx}\}$ \emph{in probability}, written as $\{\underline{\tx}_n\to\underline{\tx}\}$ a.s., if probability of
$\lim_{n\to\infty}\{\underline{\tx}_n\to\underline{\tx}\}$ equals to one.

In general stochastic root finding is capable of solving nonlinear problems as well. We denote the components of vector $\underline{\tQ}$ by $\tQ_i, i = 1, 2,\dots, N_c$ and assume that the components of $\underline{\tQ}$ are statistically independent. Note that an algorithm to solve  Eq. \eqref{RF-1} may be deterministic although the deterministic algorithm is applied to a stochastic source.

The operative definition of the statistical root finding that will be in effect throughout the present work is~\cite{Kim}:
\begin{enumerate}
  \item We consider \emph{given} a simulation that generates for any $\underline{\tPhi}\in \mathbb{R}^{N_c}$ a sequence of estimators $\underline{G}_m, m=1,2,\dots$ of the function $\underline{g}: \mathbb{R}^{N_c}\to \mathbb{R}^{N_c}$ such that $\underline{G}_m(\underline{x})\xrightarrow{d} \underline{g}(\underline{x})$ as $m\to\infty$ for all $\underline{x}$ in  $\mathbb{R}^{N_c}$.
  \item We wish to \emph{find}: a zero $\underline{x}^*\in\mathbb{R}^{N_c} $ of $\underline{g}$ such that $\underline{g^*}(\underline{x}^*)=0$ assuming $\underline{x}^*$ exists.
\end{enumerate}
Stochastic root finding does not make any assumption about the set $\underline{G}_m(\underline{x})$ except that $\underline{G}_m(\underline{x})\xrightarrow{d} \underline{g}(\underline{x})$ as $m\to\infty$. $m$ measures the simulation effort.

The norm of $\underline{\tPhi}$ is the usual $\norm{\underline{\tPhi}}= \sqrt{\sum_i (\tPhi_i)^2}$, and in an iteration the sequence of random vectors $\underline{\tPhi}_k,\;k=1,2,\dots$ is said to almost certainly converge to a random vector $\underline{\tPhi}_0$ if the probability of
\begin{equation}\label{RF-2}
    \lim_{k\to\infty} \underline{\tPhi}_k=\underline{\tPhi}_0
\end{equation}
is equal to one. Another definition is the convergence in probability when for all $\varepsilon>0$ the probability of
\begin{equation}\label{RF-3}
    \lim_{k\to\infty} \norm{\underline{\tPhi}_k -\underline{\tPhi}_0}>\varepsilon
\end{equation}
equals to zero.

There are further definitions of stochastic convergence.
\begin{enumerate}
  \item \emph{Almost sure convergence.} Let $\xi_1,\xi_2,\dots,$ be an infinite sequence of random variables defined over a subset of the real numbers $\mathbb{R}$. If the probability that this sequence will converge to a given real number $a$ equals 1, then we say the original sequence of stochastic variables converges to $a$.
  \item \emph{Convergence in probability.} Let $\xi_1,\xi_2,\dots,$ be an infinite sequence of random variables defined over a subset of the real numbers $\mathbb{R}$. If there exists a real number $a$ such that
\begin{equation}\label{CLT-3}
    \lim_{\el\to\infty} P\{|\xi_\el -a|>\varepsilon\}=0 {\text{ for all }} \varepsilon>0
\end{equation}
then the sequence converges in probability to $a$.
  \item \emph{Convergence in distribution.} Given a random variable $\xi$, with a cumulative distribution function $F(x)$, let $\xi_\el$ be a sequence of random variables, each with a cumulative distribution function $F_\el(x)$, respectively. If $\lim_{\el\to\infty} F_\el(x)= F(x)$ for all $x$ where $F(x)$ is continuous, then we say that the sequence $\xi_\el$ converges to the distribution of $\xi$.
\end{enumerate}

Note that the norm of a vector is formed from its elements, which are assumed independent and identically distributed. The norm increases with the number of elements. The definitions assure that if the Monte Carlo simulation of the source tends to be exact, the solution to equation \eqref{RF-1} tends to the exact solution. If the simulated source has a standard deviation $\sigma$ then standard deviation of the solution vector is proportional not only to $\sigma$ but also to the number of elements in $\Phi_{MC}$. The convergence of the stochastic algorithm is based on the existence of a real-valued vector-vector function $\underline{f}(\underline{x})$ such that it has a minimum at $\underline{x}^*$ and is a "smooth function"~\cite{Kim}.

There are theorems, see Theorems 5.2 and 5.3 in the survey paper~\cite{Kim}, claiming that when the involved function $\underline{f}(\underline{\tPhi})$ meets specific assumptions, the iteration converges to the deterministic limit $\underline{x}^*$. The optimal convergence rate can also be given. The convergence guarantees the solution of Eq. \eqref{RF-1} to tend to a stochastic limit. The mentioned theorems lead to the conclusions that the stochastic iteration converges to a random vector of $N_c$ components, and after a large number of iterations the limit vector is normally distributed, more precisely, as iteration number $\el\to\infty$ then $\underline{\tx}_\el\to \underline{x}^*$ and
\begin{equation}\label{AppB-1}
    \sqrt{\el}\left(\underline{\tx}_\el -  \underline{x}^*\right) \xrightarrow{d} N(0,\bSig),
\end{equation}
where
\begin{equation}\label{AppB-2}
    \bSig= \mathbf{J}^{-1T}\bSig_{\underline{x}^*}\mathbf{J}^{-1}.
\end{equation}
Here $\bSig_{\underline{x}^*}$ is the covariance matrix of the limit vector, $\mathbf{J}$ is the Jacobi matrix of the deterministic vector-vector function $\underline{g}(\underline{x})$.

The iterative solution with a random source should meet the convergence criterion \eqref{Fund-5} with given $\varepsilon$. The probability of convergence depends on the standard deviation of the source term, and the number of statistically independent elements in the random vector $\underline{\tPhi}$. This concludes the deterministic root finding ($\underline{g}(\underline{x})=0$) formalism.

In the stochastic root finding, we substitute the original deterministic equation by a stochastic equation. In the iteration we have to apply matrix $\bA$ repeatedly so we look for a stochastic approximation to $\bA$. In a stochastic experiment we use $N_p +1$ random numbers  and in one experiment we generate a  random matrix of the same order as $\bA$ in the following manner.

We build up the random matrix $\tilde{\bA}$ by drawing its elements randomly.  We write every element $A_{qr}$ as $A_{qr}= S_{qr}P_{qr}$ where $0\le P_{qr}\le 1$ and
\[
\sum_{q=1}^{N_c} P_{qr}= 1.
\]
We create a unique $\tilde{\bA}$ for a given random walk. The starting value of the matrix  $\tilde{\bA}$ is the zero matrix. We generate $N_p+1$ random numbers $\xi_1,\dots, \xi_{N_p+1}$ uniformly distributed in $[0,1]$. $\xi_1$ is used to select subscript $r$, the index of the starting cell of the random walk. When $P_{qr} \le \xi_j \le P_{q+1,r}$, the neutron wanders to cell $q$ and we increase $\tilde{A}_{qr}$ by $ S_{qr}$. We repeat the procedure until the generated random numbers have been exhausted. Then element $\tilde{A}_{qr}$ for any $qr$ subscript pair may have value $k S_{qr}$ where $0\le k\le N_p$. By the end of the procedure, we have  created a random matrix of elements
\begin{equation}\label{AppB-2-a}
    \tilde{A}_{qr}= \tilde{k}_{qr}S_{qr},
\end{equation}
where the random variable $\tilde{k}_{qr}$ is the number of random transitions from cell $r$ to cell $q$ after a random walk involving $N_p$ steps. Note that in the random matrix $\mathbf{\tilde{A}}$ at most $N_p$ elements may differ from zero. Let $\tilde{\eta}_{qr}=\tilde{k}_{qr}S_{qr}$. Then
\[
\lim_{N_p\to\infty} \frac{\tilde{A}_{qr}}{N_p}= \lim_{N_p\to\infty} S_{qr} \frac{\tilde{k}_{qr}}{N_p}= S_{qr}P_{qr}=A_{qr}.
\]
We have obtained an unbiased estimation of the matrix $\bA$ in equation \eqref{Asrf-3}.

The number  of matrix elements is $N_c^2$ and the expected number of hits per cell is $N_p/N_c^2$ with uniform probabilities.  Below we investigate the statistics of the random matrix $\tilde{\bA}$ generated in the above manner.

The following estimate of $N_p$ is given from the CLT. When $\tilde{\xi}$ is a normally distributed random variable $N(E(\tilde{\xi}),\sigma)$, to fulfill
\begin{equation}\label{AppB-3}
   P\left( \frac{\sum_{k=1}^{N_p}\tilde{\xi}_k}{N_p}- E(\tilde{\xi})<\varepsilon\right)\ge 1- p_0
\end{equation}
we need a sample of
\begin{equation}\label{AppB-4}
    N_p\ge \frac{\sigma^2}{\varepsilon^2 p_0}
\end{equation}
element. As matrix $\tilde{\bA}$ has $N_c^2$ elements, we need
\begin{equation}\label{AppB-5}
     N_p\ge \frac{\sigma^2}{\varepsilon^2 p_0} N_c^2.
\end{equation}
When $N_c=10^3, \sigma=0.01, \varepsilon=10^{-2}, p_0=10^{-2}$; then
\[
N_p\ge 10^8.
\]

  During a random walk, we have created a random approximation  $\tilde{\bA}_1$ to matrix $\bA$  such that an element $A_{qr}$ is either zero or an integer times the normalizing factor $S_{qr}$. To make use of Eq. \eqref{AppB-3}, we repeat the stochastic approximation $N_h$ times and create $\tilde{\bA}_1,\dots, \tilde{\bA}_{N_h}$. When
\begin{equation}\label{AppB-R-0}
    \lim_{N_h\to\infty} \frac{\sum_{k=1}^{N_h} \tilde{\bA}_k}{N_h}\equiv \tbR \text{  and  }E(\tbR)= \bA,
\end{equation}
we say that our statistical model is unbiased. This will be the case when the random walks follow the statistics of the physical processes underlying $\tilde{\bA}$.

Now we pass on to analyze the statistics. Let $N=N_h N_p$, and $\tilde{\eta}_{qr}=S_{qr}\tilde{k}_{qr}$. Then
\begin{equation}\label{AppB-R-0a}
    E(\tilde{\eta}_{qr})= S_{qr} E(\tilde{k}_{qr})
\end{equation}
and
\begin{equation}\label{AppB-R-0b}
    D^2(\tilde{\eta}_{qr})= S^2_{qr}D^2(\tilde{k}_{qr}).
\end{equation}
Assume that the probability of a neutron getting into cell $qr$ is independent of the cell indices and equals to $p$; let the variance of $\eta_{qr}$ be $\sigma^2$. Then the random variable
\begin{equation}\label{AppB-R-0c}
    \tilde{\zeta}_{qr}= \frac{\tilde{\eta}_{qr}-N p S_{qr}}{D(\tilde{\eta}_{qr})}\equiv \frac{\eta_{qr}-N p S_{qr}}{\sigma}
\end{equation}
is normally distributed and according to the Moivre-Laplace theorem\cite{Evans}, when $N>>1$,
\begin{equation}\label{AppB-R-5}
    P\left(a\le \tilde{\zeta}_{qr} \le b\right)= \frac{1}{\sqrt{2\pi}} \int_a^b e^{-x^2/2}dx.
\end{equation}
Thus the elements of the random matrix $\tbR_k, k=1,2,\dots$ are normally distributed with mean value $A_{ij}$ and variance $S^2_{qr}\sigma^2$. Consequently, there is an error term associated with the stochastic approximation: whenever $N$ is finite, the estimated value of the matrix element differs from the actual value. This adds an additional term to the stochastic source and increases the variance of the solution.

 In practice, we always carry out finitely many calculations and the successive approximation proceeds as  $\tbR_1,\tbR_2,\dots, \tbR_k$.
Let us write
\begin{equation}\label{AppB-R-3}
   \tbR_k = \bA + \tbH_k.
\end{equation}
The iteration with a given $k$ proceeds as follows:
\begin{equation}\label{AppB-R-4}
\begin{split}
    \underline{\tPhi}_{\el+1} & = \tbR_k \underline{\tPhi}_\el + \underline{\tQ} \\
    & = (\bA + \tbH_k) \underline{\tPhi}_\el + \underline{\tQ}; \quad\el=0,1,2,\dots
\end{split}
\end{equation}
The stochastic approximation yields a new source term which is proportional to the error of the stochastic matrix. The CLT may be used to determine the statistics of the new source term.

\section{Appendix B: Stochastic finite difference method}
Several deterministic numerical methods have their respective alter ego in the stochastic formulation, e.g. stochastic versions of the finite element method~\cite{Babuska}, and the collocation~\cite{Colloc} exists. Root finding algorithms (for example the gradient method~\cite{Gudow}) of the deterministic problem are also applicable to the stochastic problem. In the present section we discuss a fully transparent numerical problem to throw light on the nature of the stochastic iteration.

As an illustration, we show a simple example, which reflects all the aspects of the problem discussed in the present work. Consider the one-group diffusion equation in one spatial direction:
\begin{equation}\label{x.1}
    \cderii{\Phi(x)}{x} -B^2 \Phi(x) + Q =0,
\end{equation}
where $Q$ is an external neutron source independent of the flux $\Phi$. The boundary condition is
\begin{equation}\label{x.2}
    \Phi(0)=\Phi(a)=0.
\end{equation}
When it is computed by MC techniques,  the source will be a random variable that depends on the spatial variable $x$. Let the random source be
\begin{equation}\label{DE-3}
    \tilde{Q}_i= 1 + \tilde{q}_i
\end{equation}
Here $\tilde{q}_i$ is a random variable with
\[
E\{\tilde{q}_i\}=0, \qquad D^2(\tilde{q}_i)=\sigma^2
\]
for all points $x_i$, $i=1,\dots, n-1$.
\subsection{Exact solution}
When $Q$ is a constant, the solution of Eq. \eqref{x.1} is
\begin{equation}\label{x.3}
    \Phi(x)= \frac{Q}{B^2} (1-\cosh(Bx)) - \frac{Q}{B^2} \frac{1-\cosh(Ba)}{\sinh(Ba)}\sinh(Bx).
\end{equation}
When $Q$ depends on $x$, expression \eqref{x.3} is not a solution, therefore, we solve Eq. \eqref{x.1} by the finite-difference method involving an iteration. The interval $[0,a]$ is divided in $n$ intervals whose lengths are $\Delta x= a/n$. Furthermore, subscript $i$ identifies quantities belonging to $x_i=i \Delta x$. The boundary conditions are given by Eq. \eqref{x.2}. The iteration formula is obviously
\begin{equation}\label{x.4a}
    \tPhi_i= \frac{\tPhi_{i+1} + \tPhi_{i-1} + \tQ_i\cdot (\Delta x)^2}{2 + B^2(\Delta x)^2},\;\; i=1,2,\dots,n-1.
\end{equation}
Here $\tQ_i=\tQ(x_i)$ and because of the source is determined by a Monte Carlo method, $\tQ_i$ is a random variable.

It is easy to see that iteration \eqref{x.4a} is convergent for $\tQ_i\equiv 1$. Since our goal is not to study the speed of the iteration, the initial guess is always the exact solution \eqref{x.3}. When the source is perturbed by Gaussian random numbers of standard deviation $\sigma$, the iteration remains convergent but it slows down as $\sigma$ increases. Therefore it is useful to apply an over-relaxation with some suitably chosen factor $\omega$, for example $\omega\approx 1.82$ for $n=40$, see below.

Henceforth we shall assume that $E(\tQ_i)=Q \equiv 1$. Thus the linearity of equation \eqref{x.4a} entails that a similar relation holds also for the limit of the iteration \eqref{x.4a}: $E(\tPhi_i)=\Phi_i^\infty$ where $\Phi_i^\infty$ is the exact solution of the equation
\begin{equation}\label{x.4b}
    \Phi_i^\infty= \frac{\Phi_{i+1}^\infty + \Phi_{i-1}^\infty + (\Delta x)^2}{2 + B^2(\Delta x)^2},\;\;i=1,2,\dots,n-1.
\end{equation}
The $(n-1)\times (n-1)$ iteration matrix $\bA$ of equation $\eqref{x.4a}$ is
\begin{equation}\label{xxx}
    \bA= \frac{1}{2+ B^2(\Delta x)^2}
    \left(
      \begin{array}{cccccc}
        0 & 1 & 0 & \dots & 0 & 0 \\
        1 & 0 & 1 & \dots & 0 & 0 \\
        0 & 1 & 0 & \dots & 0 & 0 \\
        . & . & . & . & . & . \\
        0 & 0 & 0 & \dots & 0 & 1 \\
        0 & 0 & 0 & \dots & 1 & 0 \\
      \end{array}
    \right).
\end{equation}
It is easy to derive that the eigenvalues of $\bA$ are
\begin{equation}\label{x.5a}
    \lambda_j = \frac{2\cos \theta_j}{2+ B^2(\Delta x)^2},\;\;j=1,2,\dots,n-1
\end{equation}
with
\begin{equation}\label{x.5b}
    \theta_j= \frac{j\pi}{n}.
\end{equation}
The corresponding eigenvector is
\begin{equation}\label{x.5c}
    u_{ij}= \sqrt{\frac{2}{n}}\sin(i\theta_j),\;\;i=1,2,\dots,n-1
\end{equation}
as it can be verified by direct substitution. Let $\underline{u}_j$ be a vector with $u_{ij}$ as components. The normalization according to Eq. \eqref{x.5c} corresponds to $\norm{\underline{u}_j}=1$.
Using eigenvalues $\lambda_j$ and eigenvectors $\underline{u}_j$, we are able to solve the finite difference equations \eqref{x.4a} analytically. From the column vectors $\underline{u}_j$ we build matrix $\bU$. Since $\bA$ is symmetric, $\bU$ is a unitary matrix: $\bU^{-1}= \bU^T$. Henceforth superscript $T$ means transposed (matrix).

Let $\underline{\tQ}$ and $\underline{\tPhi}$ vectors having $\tQ_i$ and $\tPhi_i$ for components ($i=1,2,\dots,n-1$). It follows from elementary matrix algebra that the solution of the set of equations \eqref{x.4a} is
\begin{equation}\label{x.6}
    \underline{\tPhi} = \frac{(\Delta x)^2}{2+ B^2(\Delta x)^2}(\bE - \bA)^{-1} \underline{\tQ} = \frac{(\Delta x)^2}{2+ B^2(\Delta x)^2}\bU \left<\frac{1}{1-\lambda}\right> \bU^T \underline{\tQ},
\end{equation}
where $\bE$ is the unit matrix, $\left<\right>$ stands for a diagonal matrix consisting of elements within the acute brackets. In our case the $\lambda$ quantities are given in Eq. \eqref{x.5a}. Let $\underline{e}$ be a vector with all components equal to unity. We get the solution of Eq. \eqref{x.4b} by substituting $\underline{e}$ for $\underline{Q}$:
\begin{equation}\label{x.7}
    \underline{\Psi} = \frac{(\Delta x)^2}{2+ B^2(\Delta x)^2} \bU \left<\frac{1}{1-\lambda}\right> \bU^T \underline{e}.
\end{equation}
After some algebra, we obtain the following formula for the $i$-th component of vector $\underline{\Psi}$:
\begin{equation}\label{x.p}
    \Psi_i= \frac{1}{n} \frac{(\Delta x)^2}{2+ B^2(\Delta x)^2}\sum_{j=1}^{n-1} \frac{1-(-1)^j}{\tan\left(\frac{j\pi}{2n} \right)} \cdot \frac{\sin(i\theta_j)}{1-\lambda_j},\;\; i=1,2,\dots,n-1.
\end{equation}
Note that
\begin{equation*}
    \lim_{n\to\infty}\Psi_i= \Phi(x_i).
\end{equation*}
Below we investigate two iteration strategies.
\begin{enumerate}
  \item $\underline{\tQ}$ is perturbed in the first iteration step $\el=1$ and the iteration \eqref{x.4a} is followed until until convergence with $\underline{\tQ}$ left unchanged. In Appendix B, we call this a "history". The total number of followed histories is $M$, the flux is estimated via the average of the solutions obtained from $M$ converged iterations. The standard deviation of the average is estimated by means of the empirical variance.
  \item $\underline{\tQ}$ is recalculated in each iteration step \eqref{x.4a}. We shall call every such step a "history". The flux is estimated via the average of the fluxes obtained for histories. The number of histories are multiples of some integer $N$ whose values are $\el=mN$, $m=1,2,\dots,M$. The standard deviation of the average is estimated by means of the empirical variance of the averaged quantities.
\end{enumerate}
\subsection{Solution with given random source}
Equation \eqref{x.5a} lends itself to studying the iteration.
\[
\lambda= \max_j \lambda_j<1
\]
thus the optimum value of the over-relaxation factor is
\[
\omega=\frac{2}{2+ \sqrt{1-\lambda^2}}.
\]
It follows from the definitions formulated above, that the source vector can be written in the form
\[
\underline{\tQ}= \underline{e} + \underline{\tq}
\]
where we have for all components of $\underline{\tq}$ that
\[
E\{\tq_i\}=0\;\;\text{  and  } D^2(\tq_i)=\sigma^2.
\]

\begin{figure}
  \includegraphics[width=10cm]{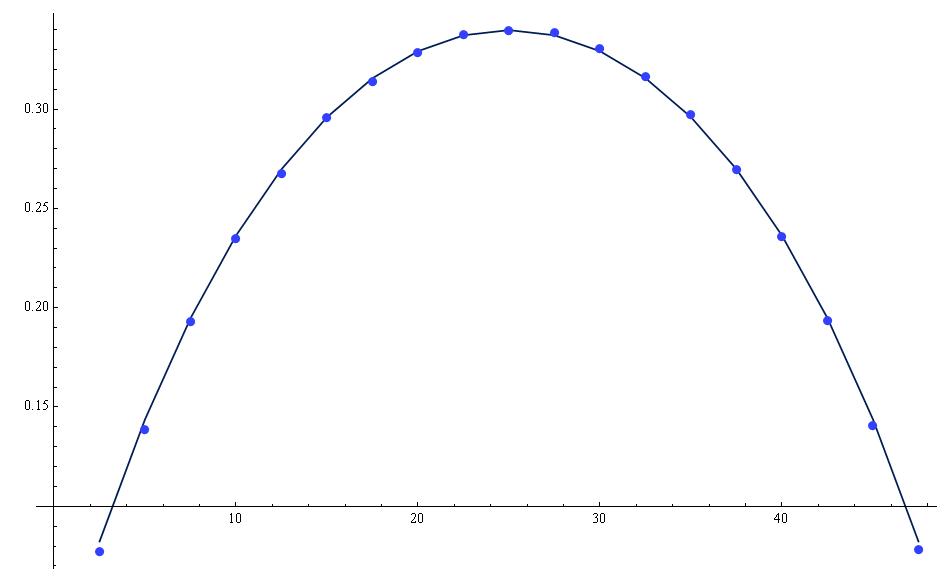}\\
  \caption{Comparison of the standard deviations given by equations \eqref{a.11} (continuous line) and equation  \eqref{a.12} (dots)} \label{F-Sz1}
\end{figure}

Below we investigate the two iteration strategies formulated earlier. The solution of the finite difference equations with a given random source $\underline{e} + \underline{\tq}$ is:
\begin{equation}\label{Sol-1}
    \underline{\tPhi}= \bA \underline{\tPhi} + \frac{(\Delta x)^2}{2+ B^2(\Delta x)^2}\underline{\tQ}.
\end{equation}
Taking the expectation values of both sides we obtain
\begin{equation}\label{a.8}
\underline{\Psi}= \bA \underline{\Psi} +  \frac{(\Delta x)^2}{2+ B^2(\Delta x)^2}\underline{e}.
\end{equation}
We introduce the deviation from the expectation values as
\begin{equation}\label{a.9}
    \underline{\tilde{\phi}}_\el = \underline{\tPhi}_\el - \underline{\Psi},
\end{equation}
and note that
\[
E(\underline{\tilde{\phi}}_\el)=0,
\]
for all $\el$. Subtracting Eq. \eqref{a.8} from Eq. \eqref{Sol-1} we find
\begin{equation*}
    \underline{\tilde{\phi}}_{\el+1} = \bA \underline{\tilde{\phi}}_{\el} + \frac{(\Delta x)^2}{2+ B^2(\Delta x)^2} \underline{\tq}.
\end{equation*}
This converges to
\begin{equation}\label{a.10}
    \underline{\tphi}= \lim_{\el\to \infty}\underline{\tphi}_\el= \frac{(\Delta x)^2}{2+ B^2(\Delta x)^2} (\bE - \bA)^{-1} \underline{\tq}.
\end{equation}
The expectation of $\underline{\tilde{\phi}}$ is the zero vector, consequently the covariance matrix is:
\begin{equation}\label{a.10-a}
    E\{\underline{\tphi} \underline{\tphi}^T\}= \left[\frac{(\Delta x)^2}{2+ B^2(\Delta x)^2}\right]^2(\bE - \bA)^{-1} E\{\underline{\tq} \underline{\tq}^T\} (\bE - \bA)^{-1}.
\end{equation}
The variances are
\begin{equation}\label{a.11}
    D^2(\tphi_i)= \sigma^2\left[\frac{(\Delta x)^2}{2+ B^2(\Delta x)^2}\right]^2 \sum_{j=1}^{n-1} u_{ij}^2 \frac{1}{(1-\lambda_j)^2},\;\;i=1,2\dots,n-1.
\end{equation}
As the components of vector $\underline{\tq}$ are independent and identically distributed, their variances being equal to $\sigma^2$, the covariance matrix is
\begin{equation*}
\begin{split}
    E\{\underline{\tphi} \underline{\tphi}^T\} & = \sigma^2\left[\frac{(\Delta x)^2}{2+ B^2(\Delta x)^2}\right]^2(\bE - \bA)^{-2} \\
    & =\sigma^2\left[\frac{(\Delta x)^2}{2+ B^2(\Delta x)^2}\right]^2 \bU\left<\frac{1}{(1-\lambda)^2}\right>\bU^T .
\end{split}
\end{equation*}
Now the variances are
\begin{equation}\label{a.11}
    D^2(\tphi_i)= \sigma^2\left[\frac{(\Delta x)^2}{2+ B^2(\Delta x)^2}\right]^2 \sum_{j=1}^{n-1} u_{ij}^2 \frac{1}{(1-\lambda_j)^2},\;\;i=1,2\dots,n-1.
\end{equation}
Let subscript $m$ identify the histories (m=1,2,\dots,M). The flux is estimated by the average
\begin{equation*}
    E\left(\tPhi_{Mi}\right)= \frac{1}{M}\sum_{m=1}^M \tPhi_{im}, \;\;i=1,2,\dots,n-1
\end{equation*}
is also a random variable.
As an average, it is always an unbiased estimate. The variance of the average is estimated by the empirical variance
\begin{equation}\label{a.12}
    s^2_{E(\Phi_{Mi})} = \frac{1}{M(M-1)}\sum_{m=1}^M (\tPhi_{im} - E(\tPhi_{Mi}))^2 = \frac{1}{M-1}\left[\overline{\tphi_i^2} - E(\tphi_i)^2 \right].
\end{equation}
Here
\begin{equation}\label{a.13}
    E\left(\overline{\tphi_i}^2\right) = \frac{1}{M^2} \sum_{m=1}^M \sum_{m'=1}^M cov(\tphi_{im},\tphi_{im'})= \frac{1}{M^2} \sum_{m=1}^M D^2(\tilde{\phi}_{im})+
    \frac{1}{M^2}\sum_{m=1}^M \sum_{m'\ne m} cov(\tphi_{im},\tphi_{im'}).
\end{equation}
Since iteration \eqref{x.6} converges for all $m$, for large $m$ expression $D^2(\tphi_{im})$ is independent of $m$. Furthermore the limiting values \eqref{a.10} obtained for the individual histories are independent of each other, the covariances vanish leading to
\begin{equation*}
    E\left(s^2_{\bar{\Phi}_i}\right)= \frac{1}{M-1}\left(D^2(\tphi_i)- \frac{D^2(\tphi_i)}{M}\right)=\frac{D^2(\tphi_i)}{M}.
\end{equation*}
Thus the empirical variance \eqref{a.12} is an unbiased estimate of the true variance derived from Eq. \eqref{a.11}. We illustrate this statement in Fig.  \ref{F-Sz1} that compares the standard deviations. The number of space points is $n=20$, the number of histories is $M=900$. This concludes the discussion of type I iteration.

\subsection{Iteration with recalculated random source}
When the source is perturbed independently in all iteration steps, which is the case when the source is recalculated in every iteration step by Monte Carlo, we have instead of Eq. \eqref{Sol-1}
\begin{equation}\label{a.14}
    \underline{\tPhi}_{\el+1} = \bA \underline{\tPhi}_\el + \frac{(\Delta x)^2}{2+ B^2(\Delta x)^2} \underline{\tQ}_{\el+1}.
\end{equation}
If we define the random vector $\tphi_\el$ by equation \eqref{a.9}, we get
\begin{equation*}
    \underline{\tphi}_{\el+1}= \bA \underline{\tphi}_\el + \frac{(\Delta x)^2}{2+ B^2(\Delta x)^2} \underline{\tq}_{\el+1}.
\end{equation*}
Similarly to the derivation given above, we simply obtain
\begin{equation*}
    \underline{\tphi}_{\el}= \bA^\el \underline{\tphi}_0 + \frac{(\Delta x)^2}{2+ B^2(\Delta x)^2} \sum_{\el'=0}^{\el-1} \bA^{\el'}\underline{\tq}_{\el-\el'}.
\end{equation*}
Let us introduce the notation
\begin{equation}\label{a.15a}
    \underline{\tzeta}_m = \frac{(\Delta x)^2}{2+ B^2(\Delta x)^2} \sum_{\el'=0}^{N-1} \bA^{\el'}\underline{\tq}_{mN-\el'}, m=1,2,\dots,M.
\end{equation}
We are interested in the results obtained for the iteration steps $\el=mN$:
\begin{equation}\label{a.15b}
     \underline{\tphi}_{mN}= \bA^{mN}\underline{\tphi}_{0} + \sum_{m'=0}^{m-1} \bA^{m'N} \underline{\tzeta}_{m-m'}.
\end{equation}
It follows from the above considerations that $E(\tphi_{mN})=E(\tzeta_m)=0$ for all $m$. Furthermore, $\tzeta_m$ and $\tzeta_{m'}$ are uncorrelated for $m\ne m'$. The average of estimates $\underline{\tPhi}_{mn}$ is an unbiased estimate of the vector $\Psi$ defined by Eq. \eqref{a.8}. Thus this iteration seems to be alright. There are, however, two problems:
\begin{itemize}
  \item first, it is hardly probable that the iteration will ever converge;
  \item secondly, the empirical variances will be biased estimates of the true variances.
\end{itemize}
In order to simplify the analysis of these problems, we assume that $mN$ is large enough for neglecting the term $\bA^{mN}\underline{\tPhi}_0$ in equation \eqref{a.15b}.

We study the convergence first. The leading terms in Eqs. \eqref{a.15a} and \eqref{a.15b} are
\begin{equation}\label{a.16}
    \underline{\tphi}_{mN}= \underline{\tzeta}_m + \bA^N \underline{\tzeta}_{m-1} + \dots = \frac{(\Delta x)^2}{2+ B^2(\Delta x)^2} \left(\underline{\tq}_{mN} + \bA \underline{\tq}_{mN-1}\right)+\dots
\end{equation}
The convergence check is based on
\begin{equation*}
    \Delta \underline{\tphi}_m= \underline{\tphi}_{mN} - \underline{\tphi}_{(m-1)N}= \frac{(\Delta x)^2}{2+ B^2(\Delta x)^2} \left(\underline{\tq}_{mN} - \underline{\tq}_{(m-1)N}+\dots\right) + \dots
\end{equation*}
To put it explicitly, the usual convergence criterion is
\begin{equation}\label{a.17}
   CONV= \max_{i}
   \abs{\frac{\tPhi_{mN,i} -\tPhi_{(m-1)N,i}}{\tPhi_{mN,i}}}
   \approx \frac{(\Delta x)^2}{2+ B^2(\Delta x)^2}
   \frac{\abs{\tq_{mN,i} -\tq_{(m-1)N,i}}}{\tPhi_{mN,i}}<\varepsilon,
\end{equation}
for some small positive $\varepsilon$, typically $\varepsilon=10^{-5}$. Although $\tPhi_{mN,i}$ is a random variable, it is an acceptable approximation to replace it by its expectation value $\Psi_i$. This, the convergence is determined by the difference $\abs{\tq_{mN,i}-\tq_{(m-1)N,i}}$ whose expectation value is $2\sigma/\sqrt{\pi}$ as it can be simply derived. Thus
\begin{equation}\label{a.18}
    CONV\approx \frac{(\Delta x)^2}{2+ B^2(\Delta x)^2} \frac{2\sigma}{\Psi_i\sqrt{\pi}}
\end{equation}
with a high probability. $CONV$ is independent of the iteration subscript $mN$. Consequently, the idea of convergence loses sense for this kind of iteration: $CONV$ will fluctuate near a limit given by Eq. \eqref{a.18}. When $CONV<\varepsilon$ the iteration stops but the opposite is also possible: $CONV$ fluctuates around the limit but the iteration will never converge, or it converges only by chance.

\begin{figure}
  \includegraphics[width=10cm]{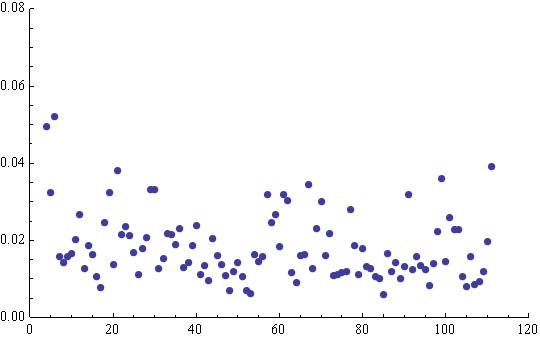}\\
  \caption{Progress of $CONV$ defined by Eq. \eqref{a.17} in the course of the iteration, $n=21$ }\label{F-Sz2}
\end{figure}

\begin{figure}
  \includegraphics[width=10cm]{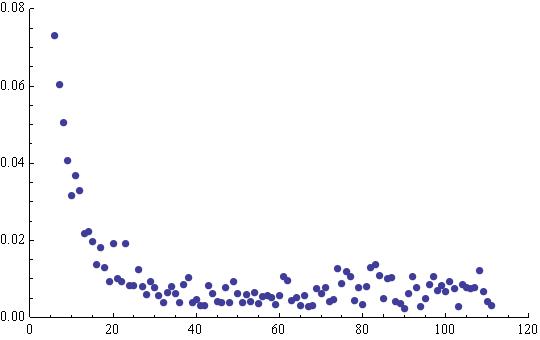}\\
  \caption{Progress of $CONV$ defined by Eq. \eqref{a.17} in the course of the iteration, $n=101$ }\label{F-Sz3}
\end{figure}

When higher powers of $\bA$ are retained in Eq. \eqref{a.16}, the formulae will change but not our conclusion. We illustrate this by Fig. \ref{F-Sz2}. To illustrate the dependence of the limit on $\Delta x$, we show the results with $n=21$ points on Fig. \ref{F-Sz2}, and with $n=101, N=9$ points on Fig. \ref{F-Sz3}.

Now we pass on to the analysis of the empirical variance. The average of the selected iterates can be approximated as follows, cf. Eq. \eqref{a.15b}:
\begin{equation}\label{a.19}
    E(\underline{\tphi})= \frac{\sum_{m=1}^M \underline{\tphi}_{mN}}{M}\cong \frac{1}{M} \sum_{m=1}^M \sum_{m'=0}^{m-1} \bA^{m'N}\underline{\tzeta}_{m-m'}.
\end{equation}
By definition, the empirical covariance matrix is
\begin{equation}\label{a.20}
    \bC^{emp}= \frac{\sum_{m=1}^M (\underline{\tphi}_{mN} - E(\underline{\tphi}))(\underline{\tphi}_{mN}-E(\underline{\tphi}) )^T}{M-1}.
\end{equation}
We have seen in Eqs. \eqref{a.12} and \eqref{a.13} that this formula gives the true variances and covariances only if the averaged quantities are uncorrelated. Therefore, we have to study the covariance between them. We need for this the following covariance matrix, see Eq. \eqref{a.15a}:
\begin{equation*}
    E(\underline{\tzeta}_m \underline{\tzeta}_m^T)= \left[\frac{(\Delta x)^2}{2+ B^2(\Delta x)^2}\right]^2 \sum_{\el'=0}^{N-1} \sum_{l''=0}^{N-1} \bA^{\el'} E(\underline{\tq}_{mN-\el'} \underline{\tq}_{mN-\el''}^T)\bA^{\el''}.
\end{equation*}
The $\underline{\tq}_\el$ vectors generated in different iteration steps are independent unless $\el'=\el''$. In the latter case, their covariance matrix is $\sigma^2\bE$. Taking this into account, we obtain
\begin{equation}\label{a.21}
\begin{split}
    E(\underline{\tzeta} \underline{\tzeta}^T) & = \sigma^2 \left[\frac{(\Delta x)^2}{2+ B^2(\Delta x)^2}\right]^2 \sum_{\el=0}^{N-1} \bA^{2\el}\\
     & = \sigma^2 \left[\frac{(\Delta x)^2}{2+ B^2(\Delta x)^2}\right]^2 (\bE- \bA^2)^{-1}(\bE-\bA^{2N}).
\end{split}
\end{equation}
Note that the result is independent of the subscript $m$.

\begin{figure}
  \includegraphics[width=10cm]{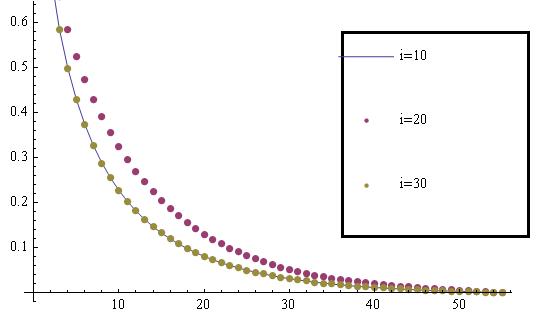}\\
  \caption{Correlation coefficients as functions of the subscript differences $k$ at positions $i=10, 20,$ and $30$}\label{F-SzXX}
\end{figure}

The terms averaged in equation \eqref{a.19} are correlated. In order to show this, we study the cross-covariances
\begin{equation*}
    \bC_{k'k''}= E(\underline{\tphi}_{(m-k')N}\underline{\tphi}_{(m-k'')N}^T)= \sum_{m'=0}^{m-k'-1} \sum_{m''=0}^{m-k''-1} \bA^{m'N} E(\underline{\tzeta}_{m-k'-m'}\underline{\tzeta}_{m-k''-m''}^T)\bA^{m''N}
\end{equation*}
for $k'$ and $k''=0,1,\dots,m-1$. According to equation \eqref{a.15a}, $\underline{\tzeta}_{m-k'-m'}$ and $\underline{\tzeta}_{m-k''-m''}$ are independent unless $m-k'-m'=m-k''-m''$. We have to distinguish two cases: $k''\le k'$ and $k'\le k''$. In the former case we can find an $m''$ for every $m'$ satisfying the condition of dependence. Thus
\begin{equation*}
    \bC_{k'k''}= \sigma^2 \left[\frac{(\Delta x)^2}{2+ B^2(\Delta x)^2}\right]^2 (\bE-\bA^2)^{-1}(\bE -\bA^{2N}) \sum_{m'=0}^{m-k'-1} \bA^{(k+2m')N}
\end{equation*}
cf. equation \eqref{a.21}. It is possible to give the matrix series in closed form, and we obtain after some elementary algebra that
\begin{equation*}
    \bC_{k'k''}= \sigma^2 \left[\frac{(\Delta x)^2}{2+ B^2(\Delta x)^2}\right]^2 (\bE-\bA^2)^{-1}\bA^{(k'-k'')N}(\bE -\bA^{[2m-2(k'-k'')]N}).
\end{equation*}
When $k'\le k''$, an analogue derivation leads to
\begin{equation*}
    \bC_{k'k''}= \sigma^2 \left[\frac{(\Delta x)^2}{2+ B^2(\Delta x)^2}\right]^2 (\bE-\bA^2)^{-1}\bA^{(k''-k')N}(\bE -\bA^{[2m-2(k''-k')]N}).
\end{equation*}
We may unite the last two formulae in
\begin{equation*}
    \bC_{k'k''}= \sigma^2 \left[\frac{(\Delta x)^2}{2+ B^2(\Delta x)^2}\right]^2 (\bE-\bA^2)^{-1}\bA^{|k''-k'|N}(\bE -\bA^{[2m-2|k''-k'|]N}).
\end{equation*}
Since all matrices commute here, we may write
\begin{equation}\label{a.21-a}
    \bC_{k'k''}=\bU \left<\mu_{k'k''}\right>\bU^T
\end{equation}
with
\begin{equation}\label{a.21-b}
    \mu_{k'k'',j}= \sigma^2 \left[\frac{(\Delta x)^2}{2+ B^2(\Delta x)^2}\right]^2  \lambda_j^{|k'-k''|N} \frac{1-\lambda_j^{(m-|k'-k''|)2N}}{1-\lambda_j^2},
\end{equation}
where $\lambda_j$ is one of the eigenvalues of $\bA$, cf. Eq. \eqref{x.5a}, the elements of matrix $\bU$ are given in Eq. \eqref{x.5c}. We recall that the element $(i',i'')$ of matrix $\bC_{k'k''}$ is the covariance of the flux iterates $\underline{\tPhi}_{i',(m-k')N}$ and $\underline{\tPhi}_{i'',(m-k'')N}$. Thus
\begin{equation}\label{a.22}
    cov(\tPhi_{i',(m-k')N},\tPhi_{i'',(m-k'')N})= \sum_{j=1}^{n-1} u_{i'j}u_{i''j}\mu_{k',k'',j}.
\end{equation}
We are interested only in this covariance for $i'=i''=i$:
\begin{equation*}
    cov(\tPhi_{i,(m-k')N},\tPhi_{i,(m-k'')N})= \sum_{j=1}^{n-1} u_{ij}^2 \mu_{k',k'',j}.
\end{equation*}
We characterize the correlations by the quantity $c_{ik}/c_{i0}$, where
\begin{equation*}
    c_{ik}= cov(\tPhi_{i,mN},\tPhi_{i,(m-k)N})=\sum_{j=1}^{n-1} u_{ij}^2 \mu_{0k,j},
\end{equation*}
with $k=0,1,2,\dots,m-1.$
Figure \ref{F-SzXX} shows the correlation coefficient $c_{ik}/c_{i0}$ as a function of $k$. The number of points is $n=40$, thus points $i=10$ and $i=30$ are symmetric in relation to the mind-interval.  It is clear that the correlation is rather high even between far away terms averaged in Eq. \eqref{a.19}. Correlation about $0.2$ can be observed between points as distant as 10-15 points.

\begin{figure}
  \includegraphics[width=10cm]{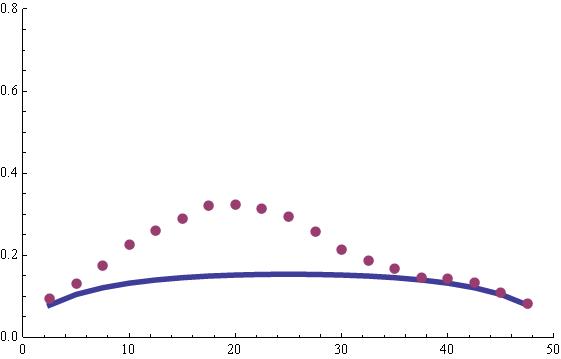}\\
  \caption{Theoretical (continuous line) and empirical (dots) standard deviations ($n=20$)}\label{F-SzAA}
\end{figure}

\begin{figure}
  \includegraphics[width=10cm]{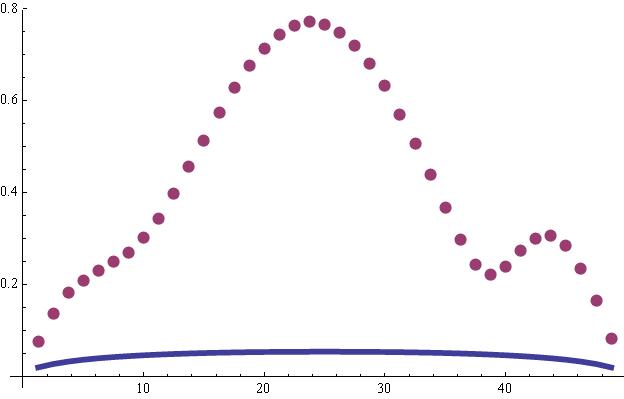}\\
  \caption{Theoretical (continuous line) and empirical (dots) standard deviations ($n=40$)}\label{F-SzBB}
\end{figure}

Although we have seen that the iteration \eqref{a.14} does not necessarily converge, we assume the opposite for simplicity. Let $M$ stand for that subscript $m$ for which the condition $\eqref{a.17}$ is satisfied. Then we have from Eq. \eqref{a.22}
\begin{equation}\label{a.23}
    D^2(\tPhi_{i,MN}) \cong \sigma^2 \left[\frac{(\Delta x)^2}{2+ B^2(\Delta x)^2}\right]^2 \sum_{j=1}^{n-1} \frac{u_{ij}^2}{1-\lambda_j^2},
\end{equation}
$i=1,2,\dots,n-1é$. This approximation holds in the limit $MN\to\infty$. On the other hand, the empirical variance \eqref{a.20} yields
\begin{equation}\label{a.24}
    D^2(\tPhi_{i,MN})=\left[\bC^{emp}\right]_{ii},\;\;i=1,2,\dots,n-1.
\end{equation}
If the empirical variance \eqref{a.24} is an unbiased estimate of the theoretical variance \eqref{a.23}, the former should stochastically converge to the latter. This is by no means the case. We illustrate this statement by figures \ref{F-SzXX}, \ref{F-SzAA}, and \ref{F-SzBB}.

\begin{figure}
  \includegraphics[width=10cm]{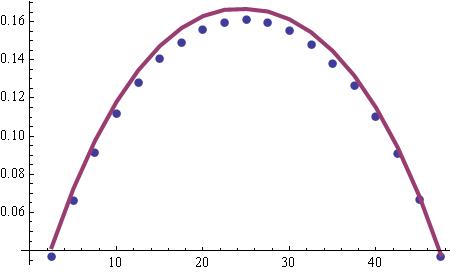}\\
  \caption{Theoretical (continuous line) and empirical (dots) standard deviations for cycle length reduced to $10^5$}\label{F-Sz5b}
\end{figure}

Finally we assess the effect of the random number generator. Every computer produces quasi-random numbers, which repeat themselves cyclically. The cycle length is usually rather large, in MCNP5 it is about $10^{19}$. The computation may surpass the capabilities of the random number generator. If so, the averaged quantities become correlated leading to biased empirical variances. In our model, the total number of the random number calls did not exceed the cycle length. In order to get an idea of the error caused by exhausting the cycle, we have artificially reduced the cycle length to $10^5$. Figure \ref{F-Sz5b} shows the resulting standard deviations: the error is nearly $4\;$\%.
\pagebreak

\pagebreak
\bibliographystyle{amsplain}

\end{document}